\documentclass{article}
\usepackage{spconf,amsmath,graphicx,hyperref}
\usepackage{booktabs}
\usepackage{tabularx}
\usepackage{makecell}
\usepackage{enumitem}
\usepackage{arydshln}
\usepackage{multirow}
\usepackage{array}
\usepackage{indentfirst}
\usepackage{tcolorbox}
\tcbset{colback=gray!5, colframe=gray!40, boxrule=0.5pt, arc=2pt, left=4pt,right=4pt,top=4pt,bottom=4pt}

\newcolumntype{L}{>{\raggedright\arraybackslash}X}

\title{Benchmarking Humans and Machines on Complex \\ Multilingual Speech Understanding Tasks}
 \name{Sai Samrat Kankanala\thanks{This work was performed with grants received from the Ministry of Education, India.}, Ram Chandra, Sriram Ganapathy}
	\address{LEAP Lab, Electrical Engineering\\
	Indian Institute of Science, Bangalore, India, 560012.\\
    \{saisamratk, ramchandra, sriramg\}@iisc.ac.in}
\begin{document}
\ninept
\maketitle
\begin{abstract}
Auditory attention and selective phase-locking are central to human speech understanding in complex acoustic scenes and cocktail party settings, yet these capabilities in multilingual subjects remain poorly understood. While machine understanding of natural speech has advanced in recent years, questions persist about comprehension of overlapped and mixed-channel speech. We propose a systematic paradigm for studying humans and machines in speech question-answering tasks in multilingual settings with clean and mixed-channel speech. For human listeners, selective attention to a target speaker was significantly better in their native language (L1) than in their second language (L2). For machine listening, speech-based large language models (LLMs) match or exceed human performance in clean, single-speaker conditions but often struggle to selectively attend in two-speaker settings. These results reveal a key divergence: humans rely on attentional cues that are more streamlined in their native language, whereas LLMs default to parallel information extraction which exceed human skills.
% Prompt design strongly influenced outcomes: \texttt{Gemini-2.5-pro} showed increased selectivity with stricter instructions, whereas \texttt{gpt-4o-audio} remained largely insensitive. 
% These results reveal a key divergence between human and machine listening. Humans rely on intrinsic attentional bottlenecks, while LLMs default to parallel information extraction unless explicitly constrained. 
% Our findings highlight the need for architectural or training modifications to embed attentional filtering, enabling models to approximate human-like selective listening.
\end{abstract}
\section{Introduction}
\label{sec:intro}
% Introduce - explain the task, why is it probed into ?
% Keep it in context - What tasks LLMs have done better, what is the need of LLMs in this task
% what we have done to the answer the above %

Humans possess a remarkable ability to selectively attend and comprehend speech in complex acoustic environments, a phenomenon famously termed the ``Cocktail party effect" by Cherry~\cite{cherry1953some}. This capability allows humans to focus on a single speaker while simultaneously processing and filtering competing speakers as well as background conversations, music, and environmental sounds. Traditionally, the dichotic listening paradigm~\cite{westerhausen2020optimal,westerhausen2019primer} has been employed to study this selective attention mechanism, wherein different auditory stimuli are presented simultaneously to each ear. This approach has provided valuable insights into hemispheric specialization~\cite{elyamany2024top} and the neural underpinnings of auditory attention~\cite{straetmans2024neural,kankanala2024uncovering} and   decoding~\cite{o2015attentional,mirkovic2015decoding,jaeger2020decoding}. 
However, a more general and realistic configuration, closer to everyday auditory environments, entails presenting a mixed diotic speech, where speech signals from two or more speakers are combined into a single audio stream and delivered identically to both ears. This overlapping speech mixture is commonly used in understanding the complex mechanisms of human speech processing \cite{mesgarani2012selective}. 
Despite its relevance to real-world scenarios, understanding similar capabilities in multilingual settings  remains limited. 

The advent of modern artificial intelligence systems, particularly multimodal large language models~\cite{team2024gemma,ghosh2025audio,goel2025audio}, has opened new questions about machine capabilities in complex auditory comprehension. These models are capable of doing tasks like Automatic speech recognition~\cite{radford2023robust} exceptionally well, even in noisy conditions~\cite{hu2024large}. Recent efforts~\cite{patman2024speech} have also demonstrated that machines can match and in some cases exceed human performance in controlled, single-speaker scenarios for speech recognition tasks. 
However, systematic understanding of how these advanced models perform tasks like question-answering when confronted with the  mixed-channel overlapping speech  is unknown. Further, the ability to follow prompt instructions of selective attention is also a matter of interest as the models continue to make progress. 

% Current state-of-the-art LLMs \cite{jiang2023mistral7b}\cite{bai2023qwen} and ALMs  \cite{chu2024qwen2}, despite their impressive capabilities in text-based reasoning and clean audio processing, face significant limitations when processing mixed audio streams. Unlike humans, who can dynamically and selectively allocate attention between competing auditory channels and maintain comprehension of multiple simultaneous speakers, these models typically struggle with the cue based selectivity, complex signal separation and contextual integration required for effective dichotic/mixed diotic listening performance. This limitation becomes particularly pronounced when the task extends beyond simple transcription to require deeper comprehension and question-answering about the mixed audio content.

% This paper presents a comparative study of human and machine performance in mixed diotic listening tasks involving overlapping speech streams and comprehension-based question answering. We evaluate how current large language and audio-language models handle the challenge of extracting content under these conditions and benchmark their performance against human listeners. Our goal is to assess the extent to which existing architectures capture aspects of human auditory scene analysis and to highlight areas where future advances in machine auditory cognition may be needed.
\begin{figure*}[t!]
    \centering
    \includegraphics[clip, trim=2cm 2cm 1cm 4.5cm, scale=0.22]{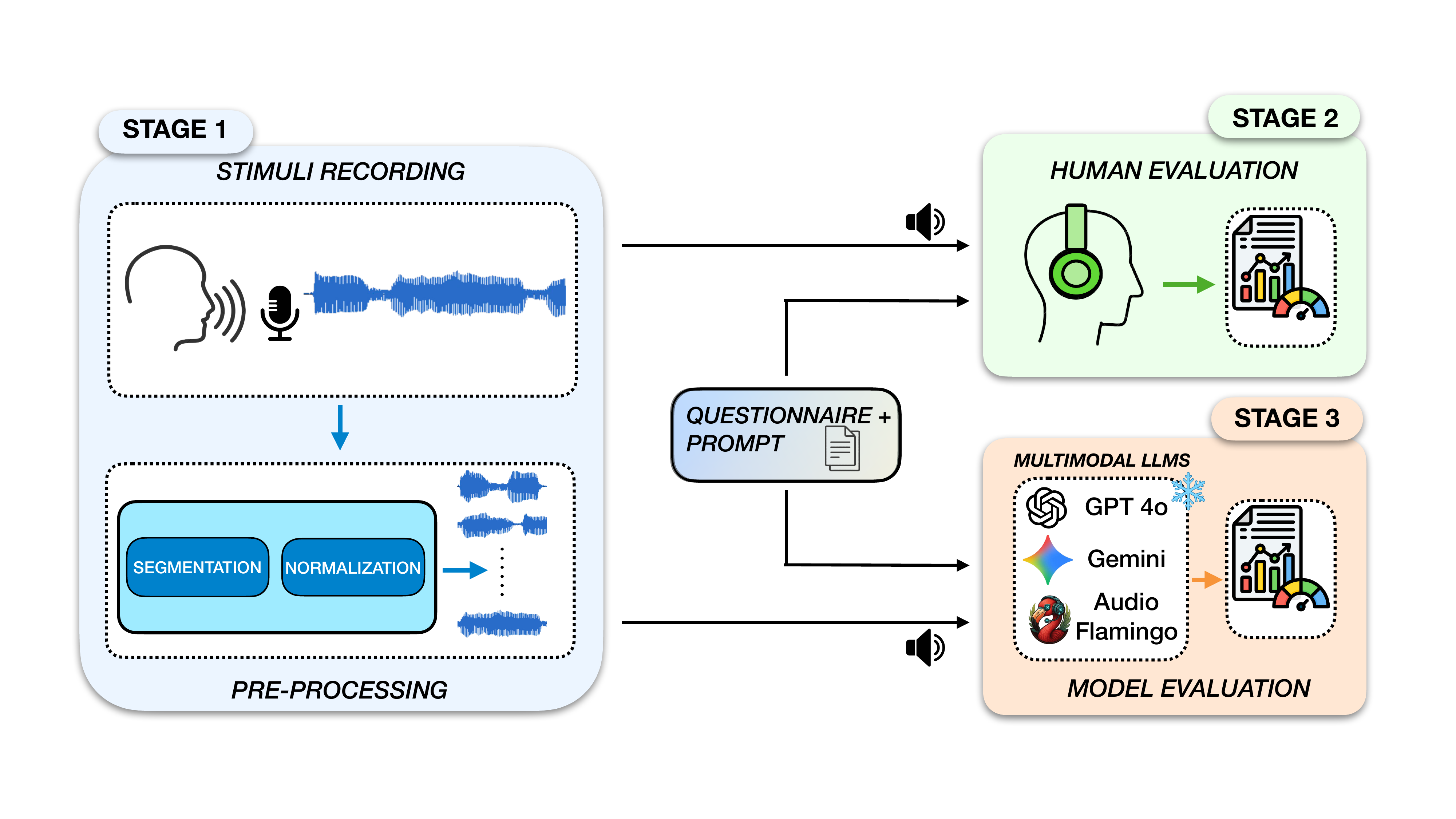}
    \vspace{-10pt}
    \caption{Schematic Illustration of the proposed framework for (i) Stimuli recording and pre-processing (Stage 1), (ii) Human evaluation (Stage 2) and (iii) Model evaluation (Stage 3).}
    \label{fig:schematic}
\end{figure*}
In this paper, we investigate the comprehension capabilities of humans and machines in complex mixed-channel speech and in multilingual settings. We have designed a speech corpus of read speech in English (Indian-English) as well as in two Indian languages (Hindi and Kannada). Using the contents of these audio files, question-answer pairs (in multi-choice form) are created and are used to probe the speech comprehension. We created two-speaker mixture files in each language separately. We recruited $40$ human subjects in our human evaluation. For model evaluation, we explored the recent \texttt{Audio-Flamingo-3} \cite{goel2025audio} model, \texttt{Gemini-2.5} (Flash and Pro models) \cite{comanici2025gemini} and the \texttt{GPT-4o} (Mini and Full) \cite{hurst2024gpt}.  
The following are the key contributions of this work:
\begin{itemize}
    \item We construct a multilingual corpus of read-speech in Indian English, Hindi, and Kannada, and derive controlled mono and mixed-channel listening materials. Further, these recordings are relatively long (about $3$ minute utterances) and capture a self-contained fictional story.  
    \item Using human-subject listening experiments, we identify the performance gap in auditory attention based selectivity in mixed-channel speech for listener's L1 (Kannada or Hindi) and L2 (Indian English).  
    \item Using speech-based LLMs, we identify that, while most of the models are good in speech question-answering in clean settings, the larger models perform significantly better in mixed-channel speech. 
    \item By comparing humans and machines, we highlight fundamental differences in selective attention mechanisms, where machine models (larger models) exceed humans significantly in their abilities to capture information simultaneously from two and three channel mixtures.
\end{itemize}

\section{Related works}
\label{sec:related_works}

\textbf{Human auditory attention:} 
Selective auditory attention has been studied extensively through controlled paradigms, with dichotic listening serving as the primary methodology~\cite{broadbent1962attention,moray1959attention}. These studies consistently reveal strong human selectivity, where listeners efficiently comprehend the attended stream while largely suppressing the unattended one~\cite{treisman1964selective}. Building on this, several works have sought to decode attended versus unattended stimuli using EEG in combination with acoustic cues~\cite{borsdorf2023multi}, semantic information~\cite{aydelott2015semantic}, or multimodal integration~\cite{wang2024self,kankanala2024uncovering}. 
However, studies on selective auditory attention in multilingual populations are  limited. 

\textbf{Large language and audio-language models:}  
The integration of speech into large language models has marked a major milestone in multimodal AI. OpenAI's Whisper demonstrated that large-scale weak supervision enables near-human robustness in automatic speech recognition across diverse languages and conditions~\cite{radford2023robust}, laying the foundation for audio-language models that combine speech understanding with language generation. Several other LLMs have also shown strong performance in tasks such as speaker diarization~\cite{yin2025speakerlm} and auditory scene analysis~\cite{goel2025audio}.  

Recent advances have enabled direct audio-to-understanding pipelines. Models such as Qwen-Audio~\cite{chu2023qwen,chu2024qwen2} and SALMONN \cite{tang2023salmonn} perform well on recognition, classification, and captioning tasks, but are restricted to relatively short audio segments (\(\approx30s\)), limiting their applicability to real-world auditory scenes involving extended dialogues and overlapping speakers. More recently, unified multimodal architectures such as GPT-4o~\cite{achiam2023gpt} and Gemini~\cite{team2023gemini} have emerged, capable of processing text, audio, and visual inputs with extended context windows. Despite these advances, direct comparisons of human and model performance in multilingual diotic listening remain scarce. Existing benchmarks  emphasize single-speaker comprehension or evaluate multi-speaker separation quality.   

\textbf{Novelty of this work:} In this work, we address this gap by creating multilingual mono- and diotic datasets in Indian English, Hindi, and Kannada, and benchmark six state-of-the-art multimodal LLMs against human listeners on comprehension-based tasks.

\section{Problem Formulation}\label{sec:problem_formulation}

We study audio question answering (AQA) in challenging multilingual and multi-speaker settings, with a particular focus on Indian languages (Kannada, Hindi) and Indian-accented English. The task is to generate an accurate answer \( y \in \mathcal{Y} \) given an audio clip \( a \in \mathcal{A} \), a textual question \( q \in \mathcal{Q} \) with candidate options \( \{o_1, o_2, o_3, o_4\} \), and an instruction prompt 
\( p \in \mathcal{P} \), where the answer must be grounded in the semantic content of the audio. We consider two categories of inputs: single-speaker mono recordings \( a_m \), which contain a single narrative voice, and mixed diotic recordings consisting of two or more overlapping speakers \( a_{\text{mix}}\), 
where distinct mono tracks are superimposed into a single diotic stream. The mixed channel audio presents a significant challenge, for humans and models, since mixture audio must be separated to enable accurate comprehension to questions in accordance with the prompt \( p \).

% Outlining the evaluation framework, model types, and challenges
The evaluation framework compares human performance with state-of-the-art end-to-end (E2E) speech language models. These models, defined as \( f_{\texttt{model}}(a, q, p) \rightarrow \hat{y} \), 
are input with the raw audio inputs together with the questions and an instruction prompt to  generate the answers. The details of the prompt are given in Table ~\ref{tab:full_prompt_structure}. 
% Beyond baseline evaluation, we explore variations of the prompt \( p \) to investigate how model performance and selectivity are affected by explicit instructions, such as focusing on a particular speaker or with emphasis on ignoring the other speaker as well. 
Humans and models are evaluated across three languages and both mono and mixed-channel speech are used as stimuli.

\section{Materials and Methods}
\label{sec:proposed_methodology}
Our methodology encompasses the usage of spoken audio data in single channel and mixed channel form as well as speech comprehension evaluation of humans and machines in these settings.
These details are provided in the following sub-sections. 
All the key steps involved in this study are outlined in Figure~\ref{fig:schematic}. 

\subsection{Speech stimuli}
\label{sec:data_collection}
In our study, we investigate long-context speech understanding in humans and machines. 
To this end, we developed our own speech stimuli as the requisite data, in the form of read speech. 
% We recorded a dataset\footnote{\textcolor{blue}{The dataset will be made publicly available at a later date.}} featuring $20$ speakers, comprising $10$ native Kannada speakers and $10$ native Hindi speakers, balanced across age and gender to ensure diversity and representativeness. 
% Both these sets of speakers were also multilingual with English as the secondary language. 
We recorded a dataset featuring $20$ speakers, comprising $10$ native Kannada speakers and $10$ native Hindi speakers, balanced across age and gender to ensure diversity and representativeness. 
Both these sets of speakers were also multilingual with English as the secondary language. 
Since open source data of this nature in multi-lingual settings was not available, we resorted to record audio data for this study. 
Each speaker narrated $20$ stories: $10$ in English, and $10$ in their native language (Kannada or Hindi), spoken with prosody and fictional style. Each story, spanning approximately $400$--$450$ words, lasted $2.5$–$3.5$ minutes in audio form, yielding a total of $20$ hours of audio data across the three languages. Our design of long-form story telling  mimics  real-world scenarios where sustained attention to longer contexts in spoken form is part of natural day-to-day human interactions. 
% Detailing question design and audio processing with silence cap rationale
Each story is paired with $10$ manually constructed, factual and context-based multi-choice questions, each with four   options. Post-recording, all audio clips were carefully trimmed to remove unwanted silence and other artifacts. Silences in the recordings were capped at $150$ milliseconds to maintain the natural flow of speech while minimizing disruptions that may impact listener attention or model performance in complex multi-speaker scenarios. Audio normalization was applied to achieve a consistent loudness level across all recordings. For mixed-channel trials, paired recordings were aligned to have approximately equal duration and energy balanced to maintain an average signal-to-interference ratio (SIR) of $0$ dB for $50$ms segments, while no explicit control was imposed on gender- or speaker-specific acoustic factors. The two/three channel mixtures were obtained by adding mono-channel audio recordings. Further, all mixtures included recordings from both male and female speakers. This design allowed the inclusion of gender-based textual prompts to guide human or machine attention.

\begin{table}[t!] % 
\centering
\caption{\label{tab:full_prompt_structure}Full prompt used for LLM evaluation.}
\footnotesize
\begin{tcolorbox}[width=0.48\textwidth] % ~half the IEEEtran column width
\textbf{Focus prompt:}  

Prompt: Please focus only on the \{MALE/FEMALE\} speaker in the audio mixture and answer the following questions.  
\vspace{1pt}\newline 
\textbf{List of Questions:}
\texttt{<Question-1,Options>, <Question-2,Options>, ... }

\vspace{0.5em}
\textbf{Instructions:}  
\begin{itemize}[leftmargin=*,nosep]
    \item Please follow the Focus prompt.  
    \item \texttt{Option}: letter (A, B, C, D) corresponding to the selected answer.    
    \item \texttt{Evidence}: one-line quote from the audio supporting the answer.  
\end{itemize}

\vspace{0.5em}
\textbf{Answer format:}  
\texttt{Question <index>}  
\texttt{Option: <A/B/C/D>}    
\texttt{Evidence: <one-line quote>}  

\vspace{0.5em}
\textbf{Input Template:}  
\texttt{<Focus prompt>, <List of questions>, <Instructions>, <Answer format>, <Audio-data>}  

\vspace{0.7em}
\textbf{Example Response:}
\texttt{Question-1 Option:B}   
\texttt{Evidence: "the boy said he was late"} 
\texttt{Question-2 Option:C}   
\texttt{Evidence: "the traffic on the route was heavy."} 
\end{tcolorbox}
\vspace{-10pt}
\end{table}

\subsection{Human Evaluation}
\label{sec:Human Baseline}
We recruited $40$ listeners  for the human evaluation study, which consisted of $20$ native Hindi speakers and $20$ native Kannada speakers, independent of the speaker pool used in the audio recordings. Participants were gender-balanced and screened to ensure normal hearing and basic language comprehension in both English and their native language (through an offline speech/text comprehension task). 
The entire listening experiment was performed in quiet conditions with a high-quality headphone.

During each session, participants were presented with mono and mixed-channel audio clips drawn from their native-language and English. 
We ensured that no listener was exposed to the same speaker or the same content more than once, throughout the listening period, thereby avoiding any familiarity bias. 
% All diotic mixtures consisted of two speakers of different genders to facilitate task performance. 

Participants listened to mono recordings or attended to a prompted gender in mixed-channel recordings before answering the questionnaire. Each mixed-channel audio had $20$ questions in the questionnaire in a random order, with $10$ questions each from two of the audio recordings in the mixture (one male and one female speaker).  
% Each diotic questionnaire contained questions pertaining to both audio streams, allowing computation of attended and unattended accuracies. 
Prior to the main sessions, participants completed a priming phase with a sample mono and a two-channel mixture with a practice questionnaires to familiarize them with the task. 
We record attended and unattended comprehension accuracy for the subsequent behavioral analysis.

\subsection{Machine Evaluation}
\label{sec:audio_qa_models}
Analogous to the human evaluation, we used the identical speech materials and prompts to evaluate the current state-of-the-art speech language models. 
In particular, we explore the application of three families of speech based LLMs:
\begin{itemize}
    \item Audio Flamingo model, version 3 \cite{ghosh2025audio}. This is open-source and relatively lightweight model of size $7$B. 
    \item Open-AI GPT \cite{hurst2024gpt} 4o series of models (both mini and full model).
    \item Gemini 2.5 \cite{team2023gemini,comanici2025gemini} series of models with flash-lite, flash and pro variants. 
\end{itemize}
The Gemini and OpenAI models are  closed models with only input-output access.   For all models, evaluation followed the same protocol as the human study: each model received an audio recording, the corresponding instruction prompt, and the questionnaire. 
The model performance was evaluated using the accuracy measure. 
The evidence responses from the model are not evaluated in this study and they are left for future analysis.

\begin{table*}[t]
\centering
\caption{\label{tab:performance_table} Comparison of human and model evaluation results (Accuracy (\%)). The human study involved $40$ participants ($20$ Hindi-native, $20$ Kannada-native) with each participant listening to $1$ mono and $1$ mixed channel trials per language, with each trial of duration $3$ minutes. The model study involved $40$ trials in each language, consisting of $20$ mono and $20$ mixed-channel trials. Further, each audio in a trial had $10$ questions. Thus, the number of trials used in evaluation is identical for both humans and machines. The color coding differentiates mono, mixture, attended, and unattended performance for both humans and models.}
\vspace{5pt}
\Large 
\renewcommand{\arraystretch}{1.2}
\resizebox{\textwidth}{!}{%
\begin{tabular}{l l l | ccc | cccccc : cccccc}
\hline 
\toprule
\multirow{4}{*}{\textbf{Method}} & 
\multirow{4}{*}{\textbf{Family}} & 
\multirow{4}{*}{\textbf{Model}} 
& \multicolumn{3}{c|}{\textbf{Mono}} 
& \multicolumn{12}{c}{\textbf{Diotic}} \\ 
\cmidrule(lr){4-6}\cmidrule(lr){7-18}
& & &  &  &  
 & \multicolumn{6}{c:}{\textbf{2 - Mixture}} 
 & \multicolumn{6}{c}{\textbf{3 - Mixture}} \\
 \cmidrule(lr){7-12}\cmidrule(lr){13-18}
 & & & English & Hindi & Kannada
 & \multicolumn{2}{c}{English} & \multicolumn{2}{c}{Hindi} & \multicolumn{2}{c}{Kannada} 
 & \multicolumn{2}{c}{English} & \multicolumn{2}{c}{Hindi} & \multicolumn{2}{c}{Kannada}\\
 \cmidrule(lr){4-6}\cmidrule(lr){7-12}\cmidrule(lr){13-18}
 & & & & & 
 & Att. & \textcolor{gray}{Unatt.} & Att. &  \textcolor{gray}{Unatt.} & Att. &  \textcolor{gray}{Unatt.} 
 & Att. &  \textcolor{gray}{Unatt.} & Att. &  \textcolor{gray}{Unatt.} & Att. &  \textcolor{gray}{Unatt.} \\
 \hline 
 \midrule
 Human    & --  & --  
 & \textbf{\textcolor{magenta}{81.3}} & \textbf{\textcolor{magenta}{95.0}}  & \textbf{\textcolor{magenta}{96.7}}
 & \textcolor{teal}{\textbf{72.5}} & \textcolor{olive}{60.4}  & \textcolor{teal}{\textbf{91.0}} & \textcolor{olive}{59.0}  & \textcolor{teal}{\textbf{80.8}}& \textcolor{olive}{45.0} 
 &  --  &  -- &  -- &  --  &  -- &  -- \\
\midrule
\midrule 
 \multirow{7}{*}{Model} 
& {AF}  & \texttt{AF-3} (7B) & \textcolor{purple}{92.0} & \textcolor{purple}{69.0} & \textcolor{purple}{50.0} & 71.5 &  \textcolor{gray}{73.5} & 34.5 &  \textcolor{gray}{31.5} & 22.0 &  \textcolor{gray}{20.5} & 62.8 &  \textcolor{gray}{60.3} & 22.5 & \textcolor{gray}{ 21.0} & 21.5 &  \textcolor{gray}{19.5}   \\
\cmidrule(lr){2-18}
& \multirow{3}{*}{Gemini} 
& \texttt{Pro 2.5} & \textcolor{purple}{90.8} & \textbf{\textcolor{purple}{100.0}} & \textbf{\textcolor{purple}{99.0}}
& \textbf{87.8} &  \textcolor{gray}{82.3} & \textbf{97.0} &  \textcolor{gray}{82.5} & \textbf{88.5} &  \textcolor{gray}{82.0}
& \textbf{79.5} &  \textcolor{gray}{73.8}  & \textbf{89.5} &  \textcolor{gray}{88.0} & \textbf{66.0} &  \textcolor{gray}{63.5} \\
& & \texttt{Flash 2.5} & \textcolor{purple}{93.0} & \textcolor{purple}{100.0} & \textcolor{purple}{98.5} 
& 81.3 &  \textcolor{gray}{76.5} & 82.5 &  \textcolor{gray}{75.5} & 78.5 &  \textcolor{gray}{76.5}
& 63.3 &  \textcolor{gray}{67.5} & 81.0 &  \textcolor{gray}{80.5} & 61.5 &  \textcolor{gray}{56.0} \\
& & \texttt{Flash-lite 2.5} & \textcolor{purple}{91.5} & \textcolor{purple}{94.0} & \textcolor{purple}{89.0} 
& 74.3 &  \textcolor{gray}{77.8} & 80.5 &  \textcolor{gray}{77.5} & 68.0 &  \textcolor{gray}{65.5} 
& 69.8 &  \textcolor{gray}{68.3} & 72.0 &  \textcolor{gray}{72.0} & 58.5 &  \textcolor{gray}{55.0} \\
\cmidrule(lr){2-18}
& \multirow{2}{*}{GPT} 
& \texttt{4o Audio} & \textbf{\textcolor{purple}{95.3}} & \textcolor{purple}{96.0} & \textcolor{purple}{95.5} 
& 85.8 &  \textcolor{gray}{84.8} & 83.5 &  \textcolor{gray}{82.0} & 73.0 &  \textcolor{gray}{70.5}
& 72.8 &  \textcolor{gray}{67.0} & 71.0 &  \textcolor{gray}{63.5} & 53.5 &  \textcolor{gray}{43.5}  \\
& & \texttt{4o Mini-audio} & \textcolor{purple}{88.8} & \textcolor{purple}{91.0} & \textcolor{purple}{84.5} 
& 77.3 &  \textcolor{gray}{76.5} & 75.5 &  \textcolor{gray}{72.0} & 56.0 &  \textcolor{gray}{50.5} 
& 47.8 &  \textcolor{gray}{48.5} & 50.5 &  \textcolor{gray}{53.5} & 28.0 &  \textcolor{gray}{28.0}  \\
\hline 
\bottomrule
\end{tabular}}

\end{table*}

\section{Results}

For benchmarking, as described in Section~\ref{sec:Human Baseline}, we calculated and reported the accuracies from the human behavioral study for both mono and diotic setups and reported in Table~\ref{tab:performance_table}. In the next set of experiments, we evaluated the LLMs~(Section \ref{sec:audio_qa_models}) using the prompt shown in Table~\ref{tab:full_prompt_structure}.
For human evaluation, we only employed two-mixture audio files, while for the model evaluation, we explored both two and three mixture audio files.  

%  For each trial $i \in \{1,\dots,N\}$, let $y_i$ be the correct option, 
% $\hat{y}_i$ the predicted option, $s_i$ the true source, $\hat{s}_i$ the predicted source, 
% and $A_i = 1$ if $\hat{y}_i \neq \texttt{CANNOT ANSWER}$ (answered), else $A_i = 0$ (abstained).  
% Then the following metrics are defined for understanding selectivity of the model.

% \begin{equation}
% \text{Answered Source Accuracy} 
%   = \dfrac{\sum_{i=1}^N A_i\,\mathbf{1}(\hat{s}_i = s_i)}{\sum_{i=1}^N A_i}.
% \end{equation}

% \begin{equation}
% \text{Abstained Source Accuracy} 
%   = \dfrac{\sum_{i=1}^N (1-A_i)\,\mathbf{1}(\hat{s}_i = s_i)}{\sum_{i=1}^N (1-A_i)}.
% \end{equation}

% \begin{equation}
% \text{Compliance score}_{\text{attended}} 
%   = \tfrac{1}{N} \sum_{i=1}^N A_i.
% \end{equation}

% \begin{equation}
% \text{Compliance score}_{\text{unattended}} 
%   = \tfrac{1}{N} \sum_{i=1}^N (1-A_i).
% \end{equation}

% \begin{equation}
% \text{Content Accuracy} 
%   = \tfrac{1}{N} \sum_{i=1}^N \mathbf{1}(\hat{y}_i = y_i).
% \end{equation}

% \textcolor{red}{The benchmarking results in Table~\ref{tab:performance_table} indicate that, with only minor exceptions, models achieve performance comparable to or exceed humans under mono-speaker conditions.}
The following are the key takeaways from the benchmarking results presented in Table~\ref{tab:performance_table}:

\begin{itemize}
    \item In human evaluations, performance in native language (Kannada/Hindi) was significantly better  than respective performance in second language (English). This is true for both the mono and the mixed-channel audio (Wilcoxon signed-rank test, $\textit{p} < 0.01$,  for all cases) (attended side). On average, this performance gap (L1-L2 gap) is $14\%, 18$\% in Hindi for mono,  2-channel audio mixture, while for Kannada the gap is $15\%, 8$\%. 
    
    \item In human evaluations, across the languages, question-answering performance on the attended audio is significantly better than for the unattended side (Wilcoxon signed-rank test, $\textit{p}<<0.01$). Note that the selective attention is instructed based on the gender of the target speaker. Thus, for each language, the performance trend is \texttt{Mono}$>$\texttt{Att.}$>$\texttt{Unatt.}  Further, the performance gap (Att.-Unatt.) is more for native languages ($32$\% in Hindi and $36$\% in Kannada) compared to English ($12$\%), indicating that it is significantly easier to perform selective attention in L1 compared to L2 (Wilcoxon signed-rank test, $\textit{p} < 0.01$).
    
    \item In model evaluations, all the closed-group models perform well on the mono conditions. The performance on Hindi and Kannada is better than those in English, largely due to the accented nature of English (Indian-English) in the stimuli.
    
    \item All models suffer a drop in performance from mono conditions to two mixture case and a further reduction in performance in the three-mixture case. Further, the performance in Hindi is significantly better than Kannada/English in all cases for the Gemini models (Wilcoxon rank-sum test, $\textit{p}<<0.01$).  In GPT models, the performance in Hindi and Indian English is similar. 
    The open-source AF models do not perform well on Hindi/Kannada data in mixed-channel settings. 
    
    \item Comparing the humans and models, the models perform significantly better than humans on Indian-English in mono as well as for both sides of the two-mixture settings (Wilcoxon rank-sum test,  $\textit{p} < 0.01$) (except for AF models in mixed-channel audio).   For Hindi and Kannada, except for the Gemini-Pro model, the human performance on the attended side is significantly better than that of the models in the two-mixture audio (Wilcoxon rank-sum test, $\textit{p} << 0.01$) . 
    
    \item Comparing humans and closed-group (Gemini/GPT) models on the unattended audio in the two-mixture setting, we observe that models perform significantly better than humans across all languages (Wilcoxon rank-sum test, $\textit{p} << 0.01$). 
\end{itemize}

These trends indicate that models are more proficient in dealing with mixed-channel audio (2-channel) compared to humans in English, as English is the L2 for the listeners. However, in the L1, except for Gemini-Pro-2.5, all models are inferior to humans in selective attention. 
Further, it is also interesting to note that the performance gap (Att.-Unatt.) is significantly lower for the models compared to humans (Wilcoxon rank-sum test, $\textit{p} < 0.01$), indicating that models are able to jointly attend to both the streams even when they are explicitly instructed to pay attention to one gender in the mixture. 
Even in the $3$-channel mixture setting, the models show limited performance difference between attended and unattended streams, highlighting a super-human capability in dealing with challenging mixed-channel audio.  
The smaller sized open-source AF models are unable to perform complex question-answering tasks in mixed-channel non-English audio. 
Finally, the Gemini-2.5-Pro exhibits remarkable speech comprehension capabilities compared to other models in our study, even for secondary languages like Kannada. 

Taken together, these results imply that selective auditory attention in humans arises from innate cognitive abilities to direct attention to specific cues that may be fundamentally better in their native language compared to a secondary language. 
Additionally, the current state-of-the-art speech large-language models show super-human speech comprehension capabilities under challenging audio conditions in multilingual settings.  However, more research efforts are required to design and develop smaller models in order to improve their abilities for such tasks.

\section{Conclusion}
In this paper, we proposed a human-machine study of speech comprehension tasks. The key probes in this study include the multilingualism as well as the ability to retrieve information  from mixed-channel overlapping speech. Towards this end, we designed a set of speech stimuli consisting of long-context stories read in a fictional style by multiple speakers. The mono-channel data were mixed to form two and three channel mixtures for our evaluation experiments.  Human participants were recruited for the evaluation, while a parallel investigation was conducted with a range of speech large language models. 
 In our evaluation, we identified several previously unknown results including (i) quantification of the performance gap in selective auditory attention between L1 and L2 for humans, (ii) multilingual performance gaps for models in mono and diotic tasks, and (iii) the lack of auditory selectivity alongside super-human abilities in large models such as Gemini-2.5-Pro. 
 As we make considerable strides in developing models for speech understanding, benchmarking humans and machines for challenging audio tasks, like the ones considered in this paper, becomes key ingredients to guide subsequent model development and deployment choices.  

%\vfill
%\pagebreak

\vfill\pagebreak

% \section{REFERENCES}
% \label{sec:refs}

\ninept
\bibliographystyle{IEEEbib}
\bibliography{refs}

\section{Compliance with ethical standards}

This study was performed in line with the principles of the Declaration of Helsinki. Approval was granted by the Ethics Committee (IHEC) of Indian Institute of Science, Bangalore ( IHEC No: 3-24072019 ), and anonymisation of participant data was ensured.

\end{document}